 \documentstyle[aps,multicol]{revtex}
\def\eei{EEI}
\def\qc{QC}

\input{epsf}
\begin{document} 
\draft 
\title{Domain walls and the conductivity of mesoscopic ferromagnets}
%\columnsep .375in 
%\twocolumn[ 
\author{Yuli Lyanda-Geller\rlap,$^{a}$ 
	I.~L.~Aleiner$^{b}$ and 
	Paul M.~Goldbart$^{a}$} 
\address{$^{a}$Department of Physics and Materials Research Laboratory,   
University of Illinois at Urbana-Champaign, Urbana, IL 61801\\ 
$^{b}$NEC Research Institute, Princeton, NJ 08540}
%----------------------------
\date{13 January 1998}
%----------------------------
\maketitle
%----------------------------
\begin{abstract} 
%----------------------------
Quantum interference phenomena in the conductivity of mesoscopic
ferromagnets are considered, particularly with regard to the effects of
geometric phases acquired by electrons propagating through regions of
spatially varying magnetization (due, e.g., to magnetic domain walls).
Weak localization and electron-electron interaction quantum corrections
to the conductivity and universal conductance fluctuations are
discussed. Experiments are proposed for multiply-connected geometries
that should reveal conductance oscillations with variations of the
profile of the magnetization.
%----------------------------
\end{abstract}
\begin{figure}[hbt]
\epsfxsize=\columnwidth
 \epsfxsize=8.0truecm
%  \centerline{\epsfbox{FIGURES/domain.EPS}} 
\centerline{\epsfbox{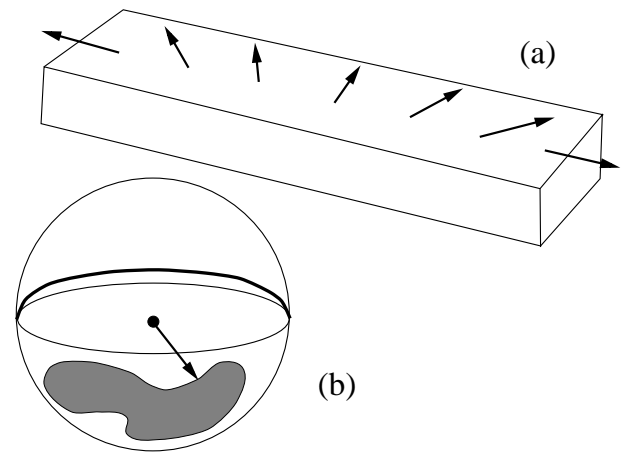}} 
 \vskip+0.4truecm
\caption{(a)~Schematic depiction of the planar variation of the 
magnetization in a simple domain wall.  (b)~Sphere of magnetization 
orientations, showing the image of the magnetization for this simple 
domain wall (the near-equatorial thick line), and for a generic 
magnetization (gray area subtending an arbitrary solid angle at 
the origin).}
\label{FIG:one}
\end{figure}

%---------------------------- 
%----------------------------
%----------------------------
\begin{multicols}{2}
\narrowtext
%----------------------------
\noindent
{\sl Introduction\/}: 
%----------------------------
In recent experiments, Giordano and Hong~\cite{Giordano2,Giordano1} 
have investigated the conductivity of ferromagnetic wires containing 
ferromagnetic domain walls, one aim being to explore the impact of 
domain walls on mesoscopic quantal phenomena.  The purpose of the this 
Letter is to present a theory of quantum effects in the conductivity of 
mesoscopic ferromagnets containing domain walls~\cite{REF:Falicov}, or 
other magnetic inhomogeneity.  Specifically, we address the
issue of the electronic resistivity due to ferromagnetic domain walls in
disordered samples, and, in particular, study weak-localization and
electron-electron interaction (EEI) quantum corrections (QC) to the
conductivity, as well as universal conductance fluctuations.  We focus
on the regime in which the wire width $L_{\perp}$ and the Fermi wave
vector $k_{\rm F}$ satisfy the conditions 
$W,L_{\perp}\gg l\gg k_{\rm F}^{-1}$, $W$ is the thickness of the domain wall.  
Our approach relies on the fact that domain walls are
characterized by an inhomogeneous magnetization  (i.e.\ a
self-consistent magnetic field that influences the electron spin), which
generates an effective spin-dependent gauge potential that is geometric
in character~\cite{REF:geo}.  This gauge potential brings with it a 
quantal geometric phase~\cite{Berry}, in much the same way that an 
electromagnetic gauge potential brings a quantal Aharonov-Bohm phase. 
The origin of this geometric phase is the variation of the magnetization 
encountered by the electron as it propagates through the medium, and its 
magnitude is determined by the solid angle swept out by the magnetization 
orientation as the electron propagates (see Fig.~\ref{FIG:one}). 

An important, and rather general, issue that we address is the \eei\
correction to conductivity in the presence of inhomogeneous magnetic
fields (either internal or external).  As we shall see, for
simply-connected geometries the inhomogeneous magnetic field suppresses
(the Diffuson channel contribution to) the \eei\ corrections to
conductivity.  Moreover, such fields do so at magnitudes lower than
those known to a influence the Diffuson channel correction due to
homogeneous magnetic fields that couple to the spin.  In addition, we
address conductance oscillations in inhomogeneous fields in
multiply-connected geometries due to \eei.  We also analyze the
influence of spin-orbit scattering by a random potential (SOI) on
Berry-phase effects in inhomogeneous fields, thus extending the
consideration of single-particle interference effects in
Ref.~\cite{REF:LSG}.  To conclude, we discuss the separation of
Berry-phase and dynamical-phase effects on conductivity of ferromagnets,
and propose feasible experiments in multiply-connected geometries.

%--------------------------
%--------------------------
It may be worth noting that our approach yields results for the
electronic resistivity in the presence of domain wall that are quite
different from those recently presented in Ref.~\cite{Fukuyama}.  In
particular, we find that a single domain wall in which the magnetization
direction varies only within a fixed plane leads to neither dephasing
nor to changes in the WL corrections to the conductivity.  By contrast,
two domain walls (or a single domain wall whose magnetization
orientation enloses a generic solid angle on the unit sphere) {\it do\/}
affect WL conductivity, the essential mechanism being a geometric phase.

The concept of geometric phase can be applied most straightforwardly when 
the relevant adiabatic condition is satisfied.  For spins this means that 
they have sufficient time complete a few precessions before the axis of 
precession changes significantly.  In the present context of the diffusive 
motion of electrons in ferromagnets, precession is caused by the (spatially 
varying) spontaneous internal magnetization ${\bf M}({\bf r})/(eg/2mc)$ 
(in which $e$, $g$ and $m$ are, respectively, the charge, $g$-factor, 
and free-electron mass) and the adiabatic condition 
becomes $M\tau\gg(\ell/L_{\rm M})(\ell/\Lambda)$, where $M$ is the 
characteristic magnitude of ${\bf M}({\bf r})$, $L_{\rm M}$ is the 
characteristic length for orientation variations of ${\bf M}({\bf r})$, 
$\tau$ is the mean free time, and $\Lambda$ is the scale of the diffusive 
path in question~\cite{REF:LSG}.  (For the case of a domain wall, 
$L_{\rm M}\sim W$.)\thinspace\ This condition is overwhelmingly 
satisfied when $M\tau\gg 1$, which is the case addressed in Ref.~\cite{Fukuyama}, 
i.e., for $d$ electrons in 
Ni or Fe.  Moreover, for $s$ electrons, for which $M$ is two orders of 
magnitude smaller, $M\tau\sim 0.1$, so that the adiabatic condition still 
holds.  Thus, the concept of the Berry phase can be readily applied to the 
calculation of the resistivity due to the domain wall.  Both $s$- and 
$d$-electron contributions to the conductivity exhibit Berry phase effects 
and, as we shall shortly see, these contributions differ in structure, and 
thus probe distinct aspects of the \qc\ to 
conductivity---interference corrections and \eei\ effects. 

%----------------------------
\noindent
{\sl Model and quantum corrections to conductivity\/}: 
%----------------------------
We now consider \qc\ to conductivity in the presense of a domain 
wall (DW). The effective one-electron Hamiltonian ${\cal H}$ of 
the ferromagnet in the presence of the DW has the form
%-----------------
\[
{\cal H}=\!\!\sum_{{\bf k},\sigma=\pm}\!\!
	\epsilon_{k}\,
	c_{{\bf k},\sigma}^{\dagger}
	c_{{\bf k},\sigma}
  	+U\!\sum_{\sigma,\sigma'}\int\!d^3 x\,
	{\bf M}({\bf x})\cdot
	c_{{\bf x},\sigma }^{\dagger}\,
	{\sigma}_{\sigma\sigma'}\,
	c_{{\bf x},\sigma'},
\]
%-----------------
where $c_{\sigma}$ and $c^{\dagger}_{\sigma}$ are annihilation and
creation operator for electrons with $z$-axis spin-projection
$\sigma=\pm$, either in position-space (${\bf x}$) or momentum-space
(${\bf k}$), $\epsilon_{k}$ ($\equiv\hbar^2k^2/2m^{*}-\epsilon_{\rm F}$)
is the single-electron energy spectrum, relative to the Fermi energy
$\epsilon_{\rm F}$, ${\bf\sigma}$ is the Pauli operator, $U$ is the
Coulomb interaction (which is self-consistently responsible for the
ferromagnetism).  To simplify our presentation we adopt units in which
$U=1$.
  
In order to calculate the interference and the \eei\ corrections to
conductivity, and also the universal conductance fluctuations, we need
to make use of the relevant Cooperon and Diffuson propagators.  By
following the standard technique of retaining the 
subset of the ladder diagrams in disorder-averaged products of
(retarded and advanced) Green functions, we readily establish the
equations obeyed by the Cooperon ($P=C$) and Diffuson ($P=D$)
propagators:
\begin{eqnarray}
&&
\Bigg[
\frac{\partial}{\partial t^{\prime}} 
+ D\left\{
i\frac{\partial}{\partial{\bf x}^{\prime}} 
+\frac{e}{c}{\rm A}_{P}^{\rm em}({\bf x}^{\prime})
\right\}^2 
+\frac{4\vert{\bf S}_{P}\vert^{2}}{3\tau_{\rm so}}
    \nonumber\\
	&&\qquad\qquad
-i{\bf M}_{\pm}({\bf x}^{\prime})\cdot{\bf S}_{\pm}\Bigg]\, P
=\frac{\delta(t-t^{\prime})\delta({\bf x}-{\bf x}^{\prime})}{\pi\nu\tau}, 
\label{EQ:CDequation}
\end{eqnarray}
where 
${\bf M}_{\pm}={\bf M}\pm{\tilde {\bf M}}$, 
${\bf A}_{C}^{\rm em}={\bf A}^{\rm em}+{\tilde{\bf A}^{\rm em}}$, 
${\bf A}_{D}^{\rm em}={\bf A}^{\rm em}-{\tilde{\bf A}^{\rm em}}$,
${\bf M}$ and ${\tilde{\bf M}}$, 
${\bf A}^{\rm em}$ and ${\tilde{\bf A}}^{\rm em}$ 
are, respectively, the internal magnetization and external 
electromagnetic vector-potential affecting two particles (1 and 2),  
${\bf S}_{\pm}=({{\bf\sigma}_1}\pm{{\bf\sigma}_2})/2$, 
${{\bf\sigma}_1}$ and ${{\bf\sigma}_2}$ are their Pauli operators, 
$S_C = S_{+}$, $S_D = S_{-}$, and $\tau_{so}$ is the SOI dephasing time.  
We note, in passing, that these equations indicate that SOI 
modifies neither the dynamical Zeeman term nor the effective 
vector-potential term, terms that were derived in Ref.~\cite{REF:LSG}.
Instead, SOI leads only to the usual dephasing
of the triplet components of the Cooperon and Diffuson propagators. 

 We now illustrate how the geometric vector potential ${\bf A}^{\rm g}$ arises due to the spatial
variation of ${\bf M}/M$. By using the adiabatic criterion we diagonalize Eq.~(\ref{EQ:CDequation}) 
in the spin space for the one-dimensional case and obtain
the equation for Diffuson in the form
\begin{equation}
\Bigg[\!D\left(i\frac{\partial}{\partial z}-A_{Sj}\!\right)^2 
+iM_{Sj}+\frac{4S}{3\tau_{\rm so}}\Bigg] D_{S,j}=
\frac{\delta(z- z^{\prime})}{\pi\nu\tau},
\label{EQ:Dsolution}
\end{equation}
where  
$ A_{Sj}=\!A^{\rm em}\!-\!{\tilde A}^{\rm em}\!+\!(1-2S)(1-\vert j\vert-j)\!(A^{\rm g}-(-1)^{j}
{\tilde A}^{\rm g})$,
	  $M_{Sj}\!=\!M+(-1)^j\!{\tilde M}$, $S$ is the net spin, 
	  and $j$ is its projection~\cite{net}.
For ${\bf M}=M(\sin{\beta}\cos{\gamma},\sin{\beta}\sin{\gamma},\cos{\gamma})$, $A^{\rm g}$ is given by 
$A^{\rm g}\!=\!(\cos{\beta (z)}\!-\!1)\partial_z\gamma/2$.	  
In the case ${\bf A}^{\rm g}={\tilde {\bf A}}^{{\rm g}}$ (${\bf M}={\tilde {\bf M}}$), which is
relevant to the interference and \eei\ correction to the conductivity,
the variation of ${\bf M}$ results in a geometric vector-potential that
enters the expression for the triplet component ($S=1$) of the Cooperon
(Diffusion) propagator for $j=\pm 1$.  In the Diffuson propagator, the
$j=\pm 1$ components are also affected by the dynamical phase arising
from the Zeeman effect of the magnetization [$M$-term in
Eq.~\ref{EQ:Dsolution}], which exists even in the case of uniform
${\bf M}/M$. Furthermore, all $S=1$ components of the
Diffuson propagator are affected by isotropic SOI dephasing,
whereas the singlet component ($S=0$) is not influenced by any of the
spin interactions.  In the Cooperon
propagator, the dynamical Zeeman phase does not affect the $j=\pm 1$
Cooperon components $C_{1,\pm}$ ($M={\tilde M}$), but these components are influenced by the 
${\bf A}^{\rm g}$ and SOI.  Instead, the Zeeman dynamical 
phase mixes $C_{00}$ with the $C_{1,0}$ component.

We now turn to the WL correction to the conductivity, which is
determined by the Cooperon propagator. In the absence of SOI, the WL
correction is not affected by the Zeeman dynamical phase, owing to the
cancellation between the contributions from the $S=0$ and the
$(S,j)=(1,0)$ components.  In the presence of SOI, the Zeeman dynamical
phase leads to a positive contribution to the magnetoresistance (MR),
owing to the noncancellation of these components. We note, however, that
in the presence of a domain wall in which $\vert{\bf M}\vert$ is
uniform, the dynamical phase does not contribute to the
resistance-change because the magnitude of this contribution is the same
in the presence and absence of the domain wall. By contrast, the
geometric phase, if any there be, owes its existence entirely to the
presence of the domain wall, and the corresponding MR is determined by
the geometric vector-potential.

Let us consider the implications of the geometric vector-potential in 
various experimental settings in (1) simply- and (2) multiply-connected samples.  
(1-a)~Consider a {\it simply\/}-connected mesoscopic wire, in which the 
magnetization varies along the wire but {\it not\/} transverse to it.  
Such magnetizations are is not uncommon and, indeed, correspond to the 
one considered in Ref.~\cite{Fukuyama}.  WL trajectories are closed (in 
position space) and, therefore, for such magnetizations, enclose no 
geometric flux.  Thus, such spatial variation of the magnetization 
generates no contribution to MR. The result of the present work differs 
from that found in Ref.~\cite{Fukuyama}.  [As ${\bf M}$ varies on 
lengthscales much larger than $\ell$ in the setting of interactions with 
domain walls, disorder-averaging was performed incorrectly in 
Ref.~\cite{Fukuyama}.]\thinspace\  (1-b)~Suppose, instead, that there 
{\it is\/} some transverse variation of the magnetization.  Then the 
image of ${\bf M}$ covers some area on the unit sphere of 
${\bf M}$-orientations. In this case, WL trajectories can include 
geometric flux, and therefore there is a contribution to MR. 
Note that in two-dimensional ferromagnets, 
topological excitations 
of magnetic order, such as skyrmions (i.e.\ two-dimensional 
textures), may lead to the modulation of the conductivity
owing to the possibility of electron trajectories enclosing differing 
amounts of geometric flux. 
(2)~Now consider a {\it multiply\/}-connected mesoscopic wire, in which
${\bf M}$ varies along the wire but not transverse to it.  In this case,
multiply-connected WL trajectories enclose multiples of the effective
geometric flux threading the hole, and therefore there will be a
contribution to MR that oscillates with the flux. For example, such a
flux-producing setting would arise in a ring containing two domain
walls, arranged so that the variation ${\bf M}$-orientation around the
ring encloses a nontrivial solid angle (and, ideally, is tunable, e.g.,
via a magnetic field).
In both cases, 1 and 2, if ${\bf M}/M$ varies
transverse to the wire then there will be a relative dephasing of
formerly coherent WL trajectories, which, in the first case leads to a
negative MR, and in the second case results in the decay of the
oscillations with geometric flux.

%----------------------------
\noindent
{\sl Electron-electron interaction effects\/}: 
%---------------------------- 
We now consider the \eei\ contribution to the \qc\ to the conductivity
in the presence of inhomogeneous magnetization orientations. \qc\ to
various kinetic and thermodynamic quantities, due to \eei\ in disordered
conductors, have their origin in the enhancement of interactions between
electrons.  The main contribution to this enhancement is due to electron
diffusion, which results in an increase of the effective interaction
time, and therefore an increase of the effective interaction strength,
that is important for electrons having close momenta and energies. This
enhancement is described by \qc\ in the Diffuson channel. Various
factors that affect electronic coherence (such as inhomogeneous
magnetization) modify the range of energies and momenta that contribute
to this diffusive enhancement of \eei, and thus modify \qc.

Of the \eei\ corrections it is those of the Hartree-type via which the
geometric gauge potential manifests itself. Here we shall focus on the
\eei\ corrections to conductivity, for which the dominant contribution
comes from processes characterized by the maximum number of diffusion
poles.  Following Ref.~\cite{Alt_Ar} we have calculated the Hartree
correction to conductivity due to the interaction of electrons and holes
with total spin~1. This correction is determined by the contribution of
the three-Diffuson processes. The two-Diffuson processes, each of them
being the same order of magnitude as the three-Diffuson terms due to
vector nature of the current vertices, compensate each other (see also
Refs.~\cite{Alt_Ar,Fukurev}).

In multiply-connected disordered samples the geometric vector-potential
results in oscillations in the conductance with variation of the solid
angle subtended by ${\bf M}$, (as well as in the density of states, and,
in general, all quantities affected by the \eei\ correction in the
Diffuson channel).  At temperatures $T$ satisfying conditions $T\tau\gg 1$ 
and $L\ge L_T$ ($\equiv\sqrt{D/T}$) the oscillating contribution to
the conductance is given by
\begin{equation}
%	\delta\sigma^{{\rm osc}}\!=\!
\frac{e^2 L_T\lambda_1}{2^{3/2}\pi\hbar L_{\perp}^2} 
\!\sum\nolimits_{n=1}^{\infty}\!e^{-n\delta}\!
\left(\sin{n\delta}-\cos{n\delta}\right)\cos{n\Omega}, 
\label{EQ:wasseven}
\end{equation} 
where $\delta\equiv L/\sqrt{2}L_T$, and $\lambda_1$ (discussed in
Ref.~\cite{Alt_Ar}) is a constant describing the interaction of an
electron and hole with total spin~1.  In ferromagnetic samples,
Eq.(\ref{EQ:wasseven}) describes the contribution from $s$-electrons.
As they are characterized by $M\tau\sim 0.1$, the triplet components of
the Diffuson propagators are affected only moderately by the Zeeman
(i.e.\ magnetization-induced) dynamical phase.  For $d$-electrons such
oscillations are very small in magnitude, because $M\tau\sim 15$ and
hence the triplet Diffuson components are entirely suppressed.  We
emphasize that the oscillations described by Eq.~(\ref{EQ:wasseven})
occur provided that not only is the condition for adiabaticity
satisfied, but also the Zeeman dynamical phase does not suppress the
correlation of electrons and holes (i.e. $M/T\le 1$).  We note that when
the latter condition is fulfilled, the usual MR due to the suppression
of the triplet-channel electron-hole interaction
contribution~\cite{Lee,Alt_Z} is absent.

Thus, for magnetizations having constant magnitude but varying
direction, the geometric vector potential is seen to affect the
conductivity in the regime where a homogeneous field of the same
magnitude plays no role. This phenomenon arises not only in ferromagnets
but also in generic conductors in the presence of arbitrary magnetic
textures.  In simply-connected settings geometric-phase-induced
anomalous magnetoresistance (due to spin effects) arises in regimes in
which (spin) dynamical-phase contributions to magnetoresistance are
negligible.  In multiply-connected settings one has the striking
situation in which inhomogeneous fields result in conductance
oscillations.  By contrast, weak homogeneous fields play no role,
whereas stronger fields would lead only to anomalous magnetoresistance
(and not oscillations).  We note that orbital magnetic fields 
(which are rather strong in ferromagnets $\sim 0.1T$) do not
affect the dominant contribution to the \eei\ corrections, namely that
arising from the Diffuson channel.  As orbital magnetic fields suppress
weak localization corrections, the oscillating \eei\ contribution can
thus be distinguished from others. In addition, geometric phase effects
can be disentangled.

%----------------------------
\noindent
{\sl Universal conductance fluctuations\/}: 
%----------------------------
We now consider stochastic variations in the conductance due to changes
in the magnetization, i.e., universal conductance fluctuations
(UCF)~\cite{REF:difference}.  In order to address these UCF one can
proceed by calculating the connected correlator
$F(A_{\rm g}^{(1)}, \delta A_{\rm g})\equiv
\langle\sigma(A_{\rm g}^{(1)})\,
\sigma (A_{\rm g}^{(1)}+\delta A_{\rm g})\rangle_{\rm c}$
between conductivities $\sigma$ at two different values of the 
geometric vector-potential, say 
$A_{\rm g}^{(1)}$ and 
$A_{\rm g}^{(2)}\equiv A_{\rm g}^{(1)}+\delta A_{\rm g}$.  
Here, the angle brackets indicate an average over realizations of the
disorder.  We restrict our attention to the
regime $L_T\ll L_{\phi}$.  In this case, the contribution to the
correlator arising from fluctuations in the diffusion
coefficient~\cite{Alt_Shkl} is essential, whereas the contribution
arising from fluctuations in the density of states is negligible.  Then,
UCF in a ring with domain walls are described by the correlator
\begin{eqnarray}
&&
F(A_{\rm g}^{(1)},\delta A_{\rm g})=
({16\pi^2}/{3})
\left({e^2}/{\hbar}\right)^2
\left({L_T^2}/{L^3}\right)
\label{EQ:correl}
\\
&&\quad\times\!\!\!
% \sum_{{i=C,D\atop{\pm}}}
\sum_{\pm,i={\rm C},{\rm D}}
L_i\left[
1+2\sum\nolimits_{n=1}^{\infty}
e^{-nL/L_i} 
\cos\left(\Omega\pm\frac{1}{2}\delta\Omega\right)
\right],
\nonumber
\end{eqnarray}
where $\delta\Omega$ is the difference between the solid angles
corresponding to the two vector-potentials, $A_{\rm g}^{(1)}$ and 
$A_{\rm g}^{(2)}$, $\Omega$ is the average of these solid angles, 
and $L_{{\rm C}({\rm D})}$ is the phase-breaking length for the 
Cooperon/Diffuson. 

The implications of Eq.~(\ref{EQ:correl}) for a single realization of the 
disorder are as follows.  For simply-connected mesoscopic wires with 
{\it no\/} (spatially) transverse variations of ${\bf M}/M$, no 
electron trajectories enclose any geometric flux.  
Therefore variations in ${\bf M}/M$ lead to neither 
deterministic nor stochastic variations in the conductance. For 
simply-connected mesoscopic wires {\it with\/} transverse variations, 
flux {\it can\/} be enclosed and, therefore, there will be stochastic
variations in the conductance. Indeed, there will effectively be a new
value of the conductance after one quantum of geometric flux is added to
the sample. For multiply-connected samples with {\it no\/} transverse 
variations, there will be conductance oscillations analogous to the $h/e$ 
(as well as $h/2e$) oscillations probed by Aharonov-Bohm quantal phases, 
familiar from mesoscopic physics.  Finally, transverse variations in 
${\bf M}/M$ will produce additional stochastic
variations of the conductance, superposed on the $h/e$-like oscillatory
variations.  We note that for $d$-electrons the Diffuson contribution to
UCF is suppressed because of the Zeeman interaction, whereas the
Cooperon contribution persists. For $s$-electrons both the Cooperon and
the Diffuson correlations are equally important for the UCF. 

As for experimental realizations, 
an external magnetic field can be used to adjust the spatial pattern of 
${\bf M}/M$, and thus alter the geometric flux.  In this 
way, one should be able to probe variations in the conductance with the 
geometric flux. Topological excitations such as skyrmions 
also modulate quantum interference effects.

%----------------------------
\noindent
{\it Acknowledgments\/}: 
%---------
We thank B.~L.~Altshuler and M.~Weissmann for interesting and useful discussions, and 
N.~Giordano for informing us of his research prior to publication.  
This work was supported by 
the Department of Energy, Division of Materials Sciences, under 
Award No.~DEFG02-96ER45439 (YLG),
and the National Science Foundation through grant DMR94-24511 (PMG).
%--------------------------------------
%----------------------------
  
%----------------------------
%----------------------------
 \end{multicols}

\begin{references}  
%----------------------------
\bibitem{Giordano2}
K.\ Hong and N.\ Giordano, 
Phys.\ Rev.\ B {\bf 51}, 9855 (1995). 
\bibitem{Giordano1}
K.\ Hong and N.\ Giordano, 
Resistance of a domain wall in a thin ferromagnetic wire, preprint (1997).
\bibitem{REF:Falicov}
At the level of classical theory the conductivity of ferromagnets having 
domain walls with a thickness much smaller than the electronic 
mean free path was addressed in the Boltzmann approximation by
A.\ Cabrera and L.\ Falicov, 
Phys. Stat. Solidi (b) {\bf 61}, 539 (1974).
\bibitem{REF:geo}
The idea of seeking implications of geometric (and related) phases in the 
context of mesoscopic phenomena has been considered quite 
widely~\cite{REF:broad,REF:LSG}.  The present setting, viz., that of 
ferromagnetic wires with domain walls, may provide a viable route for 
realising such effects. 
\bibitem{Berry} 
M.\ Berry, 
Proc.\ R.\ Soc.\ London A {\bf 392}, 45 (1984).
\bibitem{REF:broad}
D.\ Loss, P.\ M.\ Goldbart and A.\ V.\ Balatsky, 
Phys.\ Rev.\ Lett.\ {\bf 65\/}, 1655 (1990);
A.\ Stern, 
Phys.\ Rev.\ Lett.\ {\bf 68\/}, 1022 (1992)
H.\ Mathur and A.\ D.\ Stone
Phys.\ Rev.\ Lett.\ {\bf 68\/}, 1964 (1992); 
A.\ G.\ Aronov and Y.\ Lyanda-Geller, 
Phys.\ Rev.\ Lett.\ {\bf 70\/}, 343  (1993).
\bibitem{REF:LSG} 
D.\ Loss, H.\ Schoeller and P.\ M.\ Goldbart, 
Phys.\ Rev. \ B {\bf 48}, 15218 (1993).  
\bibitem{Fukuyama}
G. Tatara, H. Fukuyama, Phys.\ Rev.\ Lett.\ {\bf 78}, 3773 (1997). 
\bibitem{net}
In the Cooperon propagator the net spin is defined to be the 
{\it total\/} spin of the particles whose correlation is being 
described; in the Diffuson propagator the net spin is defined 
to be the {\it difference\/} of these spins.
\bibitem{Alt_Ar}
B.\ L.\ Altshuler and A.\ G.\ Aronov, in 
{\sl Electron-Electron Interactions in Disordered Systems\/}, 
A.\ L.\ Efros and M.\ Pollak (Eds.), p.~11, 
(North-Holland, Amsterdam, 1985). 
\bibitem{Fukurev} H.\ Fukuyama, ibid., p.~155.
\bibitem{Lee}P.\ Lee, T.\ Ramakrishnan, 
Phys.\ Rev.\ B {\bf 26}, 4009 (1982). 
\bibitem{Alt_Z}
B.\ L.\ Altshuler, A.\ G.\ Aronov and A.\ Yu.\ Zyuzin, 
Solid State Comm.\ {\bf 44}, 137 (1982).
\bibitem{Alt_Shkl} 
B.\ L.\ Altshuler and B.\ I. \ Shklovskii, 
Sov.\ Phys.\ JETP {\bf 64}, 127 (1986).
\bibitem{REF:difference} 
In Ref.~\cite{Fukuyama} UCF were addressed by drawing an analogy 
with mesoscopic fluctuations due to motion of a single 
atom~\cite{Spivak}.  The present work shows that whether or 
not changes in the magnetization lead to changes in the conductance 
depends rather sensitively on the geometrical properties of the 
magnetization configuration in question.
\bibitem{Spivak} 
B.\ L.\ Altshuler and B.\ Z.\ Spivak, 
JETP Lett.\ {\bf 42}, 447 (1985); 
S.\ Feng, P.\ Lee and A.\ D.\ Stone, 
Phys.\ Rev.\ Lett.\ {\bf 56}, 1960 (1986).
%----------------------------
\end{references}
 \end{document}